\documentclass[twocolumn]{autart}    
\usepackage{graphicx}   

\usepackage{xcolor}
\usepackage{amsmath}
\usepackage{tikzscale}
\usepackage{pgfplots}
\usepackage{tikz}
\usepackage{amssymb}
\newtheorem{assumption}{Assumption}
\newtheorem{proposition}{Proposition}
\newtheorem{lemma}{Lemma}
\newtheorem{corollary}{Corollary}
\newtheorem{theorem}{Theorem}
\newtheorem{property}{Property}

\pgfplotsset{compat=1.18}

\def\cov{{\rm cov}}
\def\vec{{\rm vec}}
\def\E{\mathbb{E}}

\def\R{\mathbb{R}}
\def\endproof{\begin{flushright} \vspace{-0.5cm} $\blacksquare$ \end{flushright}}

\def\e{\epsilon}

\definecolor{cred}{rgb}{0.8, 0.25, 0.33}
\definecolor{purple}{rgb}{0.74, 0.2, 0.64}
\definecolor{cblue}{rgb}{0.16, 0.32, 0.75}

\begin{document}

\begin{frontmatter}

\title{Covariance Propagation and Stabilization for Tube-Based Stochastic MPC under Parametric and Additive Uncertainties\thanksref{footnoteinfo}} 

\thanks[footnoteinfo]{This paper was not presented at any IFAC 
meeting. Corresponding author K.~Moussa. }

\author[UPHF,INSA]{Kaouther Moussa}\ead{kaouther.moussa@uphf.fr},   
\author[UGA]{Mirko Fiacchini}\ead{mirko.fiacchini@gipsa-lab.fr}

\address[UPHF]{UPHF, CNRS, UMR 8201 - LAMIH, F-59313 Valenciennes, France}                                              
\address[INSA]{INSA Hauts-de-France, F-59313, Valenciennes, France}            
\address[UGA]{Univ. Grenoble Alpes, CNRS, Grenoble INP, GIPSA-lab, 38000 Grenoble, France}


\begin{abstract}                          This work addresses an SMPC-oriented characterization of the error covariance dynamics for linear discrete-time systems subject to both additive and parametric stochastic uncertainties that are potentially unbounded. In contrast with the standard additive-noise case, the covariance dynamics are coupled with the nominal trajectory because the parametric uncertainty acts on the full state.
Using this characterization,  the problem of control design for error covariance dynamics is addressed, providing conditions that are conservative yet more tractable compared to standard necessary and sufficient ones for the same class of systems.  Numerical results assess this covariance characterization by comparing it to the empirical covariance and illustrate the control design problem.
\end{abstract}

\end{frontmatter}
 
\section{Introduction}
The covariance control problem, addressed in the literature since the 80s, see \cite{Collins1987,Hsieh1990,Boyd1994}, aims at controlling the covariance matrix of a linear discrete-time system affected by additive stochastic noises. Recent works consider also other types of stochastic systems, for example those subject to input constraints in \cite{Bakolas2018}, those considering chance constraints in \cite{Okamoto2018} and constant random parameters in \cite{Knaup2023}.

This paper addresses the problem of covariance control from the point of view of Stochastic Model Predictive Control (SMPC) approaches, for which the exact  characterization of covariance dynamics is useful to tighten time-varying constraints, using concentration inequalities such as the Chebyshev's inequality, as used for example in \cite{FARINA2016} and \cite{Hewing2018}. Exact characterization techniques for SMPC have mainly concerned linear discrete-time dynamical systems affected by additive stochastic uncertainties
    $x_{k+1} = A x_k + B u_k + w_k$,
with $x_k, w_k \in \mathbb{R}^n, u_k \in \mathbb{R}^m$.

One of the main approaches for uncertainties handling in Model Predictive Control (MPC) is the tube-based one \cite{Langson}. It consists in separating the state into a deterministic and an uncertain component and designing a prestabilizing feedback allowing to handle the uncertainties and their effects on chance constraints in the stochastic case. This is achieved by considering $e_k = x_k - z_k$, where $z_k \in \mathbb{R}^n$ represents the nominal deterministic component following the dynamics $z_{k+1} = A z_k + B v_k$, and the affine control law $u_k = Ke_k + v_k$, in which $K \in \mathbb{R}^{m \times n}$ represents the pre-stabilizing feedback. Therefore, $e_{k+1}=(A+BK)e_k+w_k$ and, if $e_0=0$, then the expectation of $e_k$ is also null, which leads to the following covariance dynamics of $e_k$ under the assumption that the sequence of $w_k$ is i.i.d. with respect to time $k$:
\begin{equation}
    \text{cov}(e_{k+1}) = (A+BK)\text{cov}(e_{k})(A+BK)^T + W,
    \label{Cov_Additive}
\end{equation}
where $W$ is the covariance of $w_k$, i.e. $\cov(w_k)=\E[w_k w_k^T]=W$ when $\E[w_k]=0$. Thus, in the additive-noise case, stabilizing the covariance dynamics in (\ref{Cov_Additive}) amounts to choosing $K$ such that $A+BK$ is Schur, and this autonomous recursion can be used for reachability analysis and SMPC constraint tightening \cite{KofmanAUT12,Fiacchini2021,Arcari2023}. In particular, chance constraints of the form $\Pr\{x_k\in\mathcal X\}\geq 1-\delta$, with $\delta \in (0,1)$,  can be handled by constructing $k$-step reachable sets $\mathcal R_k$ for $e_k$ and imposing deterministic constraints $z_k\in\mathcal X\ominus\mathcal R_k$, where $\ominus$ is the Pontryagin set difference; avoiding the scenario generation required by randomized methods \cite{Cannon2011,Lorenzen2016,Blackmore2010,calafiore2012}. With multiplicative parametric uncertainties, however, this standard recursion is no longer valid; the error dynamics depends explicitly on the uncertain parameters and, as shown in this paper, the covariance propagation is coupled with the nominal trajectory. This makes the design of reachable sets and SMPC tightening mechanisms more involved than in the purely additive-noise case.

Multiplicative uncertainties in linear dynamics have a long history, see \cite{Boyd1994} for early references. Restricting our attention to discrete-time systems, the work \cite{Li2005} should be mentioned, where a framework is presented for the design of a feedback controller and an estimator using LMIs. The more recent work \cite{Gravell2021} presented a framework for the infinite-horizon linear quadratic regulator problem with multiplicative noise. Vectorized second-moment dynamics have also been used in identification-oriented works for linear systems with multiplicative noise, see for instance \cite{Xing2020ACC,Xing2022AUT}. Mean Square Stability (MSS) of systems with state and input stochastic uncertainties has been addressed in \cite{HOSOE2019}, where the authors derive necessary and sufficient conditions allowing to design a state feedback control to ensure MSS for the closed-loop system. Since this work is closely related to our results, related comparisons will be presented.

\subsection*{Contribution}
The main contribution of this technical note is to provide an SMPC-oriented characterization of the error covariance dynamics when both multiplicative and additive uncertainties (of stochastic nature and with potentially unbounded support) affect a discrete-time linear system. The characterization is obtained for the tube-oriented decomposition $x_k=z_k+e_k$ and shows explicitly that, unlike in the standard additive-noise case, the error covariance dynamics are coupled with the nominal trajectory through a term depending on $z_kz_k^T$. This coupling is important for SMPC, since reachable sets and constraint-tightening mechanisms cannot be obtained from an autonomous error covariance dynamics only. Furthermore, we derive a Linear Matrix Inequality (LMI)-based condition for covariance control design, which allows to stabilize the covariance dynamics, considering a decay rate minimization. In this case, the covariance dynamics involves quadratic terms of the pre-stabilizing feedback gain $K$ resulting from a Kronecker product property. Finally, we show that, although conservative, this condition is significantly more tractable (from a computational point of view) for higher dimensional systems than conditions reported in \cite{HOSOE2019}.  Future works will exploit the derived dynamics to address the above-mentioned coupling in the context of reachability analysis, and will tackle covariance optimization in order to maximize the SMPC feasiblity region. 
\subsection*{Notation}
Denote with $\mathbb{R}$ and $\mathbb{N}$, respectively, the sets of real and natural numbers. The expectation of a random variable $x$ is denoted by $\mathbb{E}[x]$. Given a random vector $v$, $\cov(v) = \mathbb{E}[(v-\mathbb{E}[v])(v-\mathbb{E}[v])^T]$ stands for the covariance of $v$, if the latter has a zero mean ($\mathbb{E}[v]=0$), then the covariance of $v$ is simply $\cov(v)=\mathbb{E}[vv^T]$. The normal distribution of mean $\mu$ and covariance matrix $\Sigma$ is denoted $\mathcal{N} \left( \mu, \Sigma\right)$. The Kronecker product is denoted by $\otimes$, $\vec(\cdot)$ stands for the vectorization operator and $\vec(\cdot)^{-1}$ stands for the inverse of the vectorization operator. Given a square matrix $A \in \mathbb{R}^{n \times n },$ with $n \in \mathbb{N}$, $\rho(A)$ and $\sigma_{max}(A)$ stand, respectively, for the spectral radius and the maximal singular value of $A$.      The notaton $\{ \cdot \}_{ms}$ stands for a multiset which is a collection of elements that allows for repetition. The multiset consisting of the eigenvalues of $A$ including their algebraic multiplicity is denoted by $\textnormal{mspec}(A)$. The zero  and identity matrices of appropriate dimensions are denoted, respectively, $0$ and $I$. Given a symmetric matrix $M$, $\lambda_{max}(M)$ denotes the largest eigenvalue of $M$ having real eigenvalues, and  $M \succ 0$ means that $M$ is positive definite. \\

\section{Problem statement}

Consider the following discrete-time linear system:
\begin{equation}
    x_{k+1} = A(p_k) x_k + B u_k + w_k,
\label{Eq:sys_dyn}
\end{equation} 
where $x_k \in \mathbb{R}^{n}$ and $u_k \in \mathbb{R}^{m}$ represent, respectively, the state and the control input. The initial state $x_0$ is assumed to be deterministic.  We denote by $p = \left( p_k\right)_{k \in \mathbb{N}}$ and $w = \left( w_k\right)_{k \in \mathbb{N}}$, $l$ and $n$-dimensional stochastic i.i.d. sequences, representing parametric and additive uncertainties, respectively. We assume that $\mathbb{E}[w_k]=0$ and $\cov(w_k) = \mathbb{E}[ (w_k-\mathbb{E}[w_k])(w_k-\mathbb{E}[w_k])^T]=\mathbb{E}[w_kw_k^T]=W$, with $W \succ 0$. Moreover, we assume that the square elements of $A(p_k)$ are all Lebesgue integrable, \textit{i.e.},
$\mathbb{E}\left[A_{i,j}(p_k)^2 \right] < \infty$ for all $\: i,j=1,\ldots,n$, where $A_{i,j}(p_k)$  denotes the $(i,j)$ entry of $A(p_k)$.

Note that this assumption is standard when studying the stability of the covariance dynamics or MSS for linear systems as mentioned in \cite{HOSOE2019}. Without loss of generality we can rewrite $A(p_k)$ as $A(p_k)=A_0+\bar{A}(p_k)$, where $\mathbb{E}\left[\bar{A}(p_k) \right] = 0$ and $A_0 = \mathbb{E}\left[A(p_k)\right]$ represents the mean of the random matrix. The pair $(A_0, B)$ is assumed stabilizable. 

    Furthermore, we assume that the elements of $\bar{A}(p_k)$ are mutually independent of the elements of $w_k$. This assumption is not restrictive, it is only considered to simplify the exact characterization of the covariance dynamics, and additional terms resulting from its non-satisfaction (that can be easily considered) do not affect the stability analysis of the covariance dynamics. Since the sequences of $p_k$ and $w_k$ are i.i.d,  then the elements of $\bar{A}(p_k)$ and those of $w_k$ are independent of  the state for the same time step $k$, meaning that $\mathbb{E}\left[ \bar{A}(p_k) x_k\right] = \mathbb{E}\left[\bar{A}(p_k)\right] \mathbb{E}\left[x_k\right]=0$  and $\mathbb{E}\left[x_kw_k^T\right]=\mathbb{E}\left[x_k\right]\mathbb{E}\left[w_k^T\right]=0$.

Given system~(\ref{Eq:sys_dyn}) and the assumptions formulated above, the problem addressed in this paper is finding an exact expression of the error covariance dynamics induced by the tube-oriented decomposition, to be employed within the stochastic MPC framework. Moreover, we are interested in studying the stability of these covariance dynamics in order to derive a synthesis condition for the pre-stabilizing feedback gain that guarantees the convergence of the error covariance.

\section{Exact covariance characterization}
Consider system~(\ref{Eq:sys_dyn}), the state $x_k$ can be expressed as the sum of a deterministic component $z_k$ and a random component $e_k$ that is 
\begin{equation}
    x_k = z_k + e_k,
\end{equation}
such that
\begin{equation}\label{eq:z}
    z_{k+1} = A_0 z_k +B v_k,
\end{equation}
with $z_0 = x_0$ and then $e_0 = 0$. From $e_k = x_k-z_k$ and by considering $u_k = Ke_k + v_k$ we have:
\begin{align}
    e_{k+1} & = x_{k+1}-z_{k+1} = (A_0+BK)e_k+\bar{A}(p_k)x_k+w_k \nonumber\\
    &=(A(p_k)+BK)e_k+\bar{A}(p_k)z_k+w_k.\label{eq:e}
\end{align}
The following standard assumption is functional to the subsequent results and is not restrictive since $(A_0, B)$ is assumed to be stabilizable, which is commonly used in standard MPC methods. 
\begin{assumption}\label{Ass:exp_stab}
The system (\ref{eq:z}) is exponentially stabilized by the control $v_k$.
\end{assumption}
In practice $v_k$ can be designed by solving a deterministic MPC problem, as usual for tube-based approaches. The following lemma shows that $\mathbb{E}[e_k]=0$ which helps in the exact characterization of the covariance dynamics presented subsequently. 
\begin{lemma}\label{Prop:1}
From $e_0 = 0$ it follows that $\mathbb{E}[e_k]=0$ for all time instants $k$.  
\end{lemma}
\paragraph*{Proof} 
The expectation of the error dynamics is  
\begin{align}
    \mathbb{E}[e_{k+1}]&=\mathbb{E}[(A(p_k)+BK)e_k+\bar{A}(p_k)z_k +w_k] \nonumber\\
    &=\mathbb{E}[(A(p_k)+BK)e_k]+\mathbb{E}[\bar{A}(p_k)z_k]+\mathbb{E}[w_k].\nonumber
\end{align}
Since $A(p_k)$ is independent of both $e_k$ and $z_k$  and the expectation of the product of two independent random variables is the product of their respective expectations \cite{Bertsekas2008} then it follows:  
\begin{equation*}
    \mathbb{E}[e_{k+1}]=\mathbb{E}[(A(p_k)+BK)]\mathbb{E}[e_k]+\mathbb{E}[\bar{A}(p_k)]z_k+\mathbb{E}[w_k].
\end{equation*}
Moreover, since $\mathbb{E}[\bar{A}(p_k)]=0$ and $\mathbb{E}[w_k]=0$, then
\begin{equation*}
\mathbb{E}[e_{k+1}]=\mathbb{E}[(A(p_k)+BK)]\mathbb{E}[e_k],
\end{equation*}
and therefore, since $e_0=0$, we have that $\mathbb{E}[e_k]=0$ for all time instants $k$. 
\vspace{-0.3cm}
\endproof

A direct implication of Lemma \ref{Prop:1} is that $\cov(e_k) = \mathbb{E}[e_k e_k^T]$. The following properties are used hereafter for the covariance exact characterization proof.

\begin{property}\label{pr:BCA} \textnormal{(Proposition 7.1.9. in \cite{Bernstein2009})}\\
Let $A \in \mathbb{R}^{n \times m}, B \in \mathbb{R}^{m \times l}$ and $C \in \mathbb{R}^{l \times k}$, then:
\begin{equation*}
  \vec (ABC)= \left(C^T \otimes A\right) \vec(B). 
  \label{Eq:property_kron}
\end{equation*}
\end{property}

\begin{property} \textnormal{(Proposition 7.1.6. in \cite{Bernstein2009})}\\
    Let $A \in \mathbb{R}^{n \times m}, B \in \mathbb{R}^{l \times k }, C \in \mathbb{R}^{m \times q}$ and $ D \in \mathbb{R}^{k \times p}  $, then:
\begin{equation*}
    \left(A \otimes B \right)\left(C \otimes D\right) = AC \otimes BD.
\end{equation*}
\label{Property_Kronecker}
\end{property}
The following proposition provides the covariance dynamics of the error, which will be used to analyze covariance convergence and synthesis conditions.
\begin{proposition} \label{th:1}
The dynamics of the error covariance related to system~(\ref{Eq:sys_dyn}) is given by the following equivalent expressions:  
\begin{align}
& \cov(e_{k+1})=(A_0+BK) \cov(e_k)(A_0+BK)^T  +W \nonumber\\
& + \vec^{-1} \left(\mathbb{E}[\bar{A}(p_k)\otimes\bar{A}(p_k)]\vec\left( \cov(e_k)+z_kz_k^T\right) \right),
\label{Eq:cov_dyn_w}
\end{align}
and 
\begin{equation}
     \epsilon_{k+1}= \Big( (A_0+BK) \otimes (A_0+BK) + C_p \Big) \epsilon_k + C_p \zeta_k + \omega ,
    \label{eq:err_cov_dynamics}
\end{equation}
where $\epsilon_k=\vec\left(\cov(e_k) \right)$, $\zeta_k = \vec \left( z_kz_k^T\right)$, $\omega = \vec \left( W\right)$ and  $C_p=\mathbb{E}[\bar{A}(p_k)\otimes\bar{A}(p_k)]$.
\end{proposition}
\paragraph*{Proof} From (\ref{eq:e}) and Lemma \ref{Prop:1}, it follows
 \begin{align*}
\cov(e_{k+1}) & =\mathbb{E}\Big[\left( \left(A_0+BK\right)e_k+\bar{A}(p_k)x_k+w_k\right) \\ 
& \hspace{0.4cm} \cdot ( \left(A_0+BK\right)e_k+ \bar{A}(p_k)x_k +w_k)^T \Big]\\ 
& = \left(A_0+BK\right) \mathbb{E}[e_ke_k^T] \left(A_0+BK\right)^T\nonumber\\
& \hspace{0.4cm} +\mathbb{E}[\bar{A}(p_k)x_kx_k^T\bar{A}(p_k)^T] + \mathbb{E}[w_kw_k^T],
\end{align*}
since $\bar{A}(p_k)$ and $w_k$ are mutually independent from $x_k$ and $e_k$. The first term is dependent on the error covariance $\cov(e_k) = \mathbb{E}[e_ke_k^T]$, while the second one, resulting from the presence of uncertain parameters, is given by
\begin{align}
&\mathbb{E}[\bar{A}(p_k)x_kx_k^T \! \bar{A}(p_k)^T \! ] \! = \! \mathbb{E}[\bar{A}(p_k)(e_k \! + \! z_k \! )(e_k \! + \! z_k \! )^T \! \bar{A}(p_k)^T  \! ] \nonumber \\
& \hspace{0.25cm} = \mathbb{E}[\bar{A}(p_k)e_ke_k^T\bar{A}(p_k)^T]+\mathbb{E}[\bar{A}(p_k)z_kz_k^T\bar{A}(p_k)^T] \nonumber \\
& \hspace{0.25cm}  + \mathbb{E}[\bar{A}(p_k)e_kz_k^T\bar{A}(p_k)^T] +\mathbb{E}[\bar{A}(p_k)z_ke_k^T\bar{A}(p_k)^T]. \label{Eq:cov_p}
\end{align}
By using Property~\ref{pr:BCA} on the different terms of (\ref{Eq:cov_p}), from the linearity of the vectorization operator, and the fact that the expectation of a matrix is the matrix of expectations, implying that the vectorization operator and the expectation can commute, we obtain:
\begin{align*}
   &\vec \left( \mathbb{E}[\bar{A}(p_k)x_kx_k^T\bar{A}(p_k)^T]\right) = \vec \Bigl( \mathbb{E}[\bar{A}(p_k)e_ke_k^T\bar{A}(p_k)^T]  \nonumber\\
    &\! + \! \mathbb{E}[\bar{A}(p_k)z_kz_k^T \!\!\bar{A}(p_k)^T] \! + \! \mathbb{E}[\bar{A}(p_k)e_kz_k^T \!\! \bar{A}(p_k)^T ] \! \nonumber\\
    & \! + \! \mathbb{E}[\bar{A}(p_k)z_ke_k^T\bar{A}(p_k)^T] \!\Bigr) \nonumber\\
&  \!  =  \! \vec \! \left( \mathbb{E}[\bar{A}(p_k)e_ke_k^T\bar{A}(p_k)^T]\right) \!  +  \! \vec \! \left(\mathbb{E}[\bar{A}(p_k)z_kz_k^T\bar{A}(p_k)^T] \right) \nonumber\\ 
&  \! +  \! \vec  \! \left( \mathbb{E}[\bar{A}(p_k)e_kz_k^T\bar{A}(p_k)^T]\right)  \! +  \! \vec  \! \left(\mathbb{E}[\bar{A}(p_k)z_ke_k^T\bar{A}(p_k)^T] \right) \! . \nonumber
\end{align*}
From Lemma~\ref{Prop:1} and Property~\ref{pr:BCA} it follows 
\begin{align}
& \vec  \! \left(\mathbb{E}[\bar{A}(p_k)e_kz_k^T\bar{A}(p_k)^T] \right)  \! =  \! \mathbb{E}\left[\vec \! \left( \bar{A}(p_k)e_kz_k^T\bar{A}(p_k)^T\right)\right] \nonumber \\
& \! = \! \mathbb{E}[\left( \! \bar{A}(p_k) \! \otimes \! \bar{A}(p_k) \right) \!\vec \!\left(e_kz_k^T \right)] \!\nonumber\\
& = \! \mathbb{E}[\left( \! \bar{A}(p_k) \! \otimes \! \bar{A}(p_k) \right) ] \vec \! \left(\mathbb{E}[e_kz_k^T ]\right) \nonumber\\
& \!= \! \mathbb{E}[\left(\bar{A}(p_k) \otimes \bar{A}(p_k) \right)] \vec \left(\mathbb{E}[e_k]z_k^T\right) = 0.\label{Eq:Proof_kron}
\end{align}
Analogous results hold for the term $\mathbb{E}[\bar{A}(p_k)z_ke_k^T\bar{A}(p_k)^T]$, and hence, following the same steps as in (\ref{Eq:Proof_kron}), one has: 
\begin{align}
&  \vec \left( \mathbb{E}[\bar{A}(p_k)x_kx_k^T\bar{A}(p_k)^T]\right) = \nonumber \\ 
&\hspace{0.5cm} \mathbb{E}[\bar{A} (p_k)\otimes \bar{A} (p_k)] \vec \left( \mathbb{E}[e_ke_k^T]\right) \nonumber \\ 
&\hspace{0.5cm}+ \mathbb{E}[\bar{A} (p_k)\otimes \bar{A} (p_k) ]\vec \left(z_kz_k^T\right) ,       \label{Eq:penultimate_proof}
\end{align}
$z_k$ being deterministic. Finally, (\ref{Eq:cov_dyn_w}) follows from  (\ref{Eq:penultimate_proof}).  By defining $\epsilon_k = \vec(\cov(e_k)) \in \R^{n^2}\!\!, \ \zeta_k = \vec((z_k z_k^T)) \in \R^{n^2}\!\!$, $ \omega_k = \vec(\cov(w_k)) \in \R^{n^2}$ and $C_p = \mathbb{E}[\bar{A}(p_k)\otimes\bar{A}(p_k)]$, equation (\ref{eq:err_cov_dynamics}) follows directly.
\endproof

Proposition~\ref{th:1} shows thereby that the covariance evolves like a linear controlled system whose dynamics is affected by uncertain parameters through the deterministic matrix $C_p = \mathbb{E}[\bar{A}(p_k)\otimes\bar{A}(p_k)]$ containing the second-order moments related to the entries of $\bar{A}(p_k)$. It extends the additive-noise covariance dynamics used, for example, in \cite{KofmanAUT12,Fiacchini2021}, to the error dynamics under stochastic parametric uncertainty. The key difference with respect to the purely additive-noise case is the presence of the term $C_p\zeta_k$, which couples the covariance dynamics with the nominal trajectory $z_k$. This point is particularly relevant for SMPC, since the reachable sets used for constraint tightening depend on the nominal trajectory in the transient phase and a careful analysis needs to be carried out in order to jointly design the deterministic MPC strategy over the mean dynamics and the reachable sets tightening.

By defining $M(K) = \left( (A_0+BK) \otimes (A_0+BK) + C_p \right)$, the following corollary provides the limit of the error covariance if the system $\epsilon_{k+1} = M(K) \epsilon_k$ is asymptotically stable and the nominal dynamics is exponentially stabilized.
\begin{corollary}\label{Cor:Stability}
Consider $M(K)$, assume that $K$ is such that $\rho\left( M(K) \right) < 1$, and let Assumption~\ref{Ass:exp_stab} hold. Then, $\zeta_k=\vec(z_kz_k^T)$ converges to zero and $\cov(e_k)$ converges to the unique stationary covariance
\begin{equation*}
\vec^{-1}\left(\left( I-M(K)\right)^{-1} \vec\left(W\right)\right).
\end{equation*}
\end{corollary}
\section{Covariance-stabilizing  controller design}
The following properties will be used to derive an LMI condition for the design of the pre-stabilizing gain $K$.
\begin{property} \textnormal{(Fact 5.12.2., page 333 in \cite{Bernstein2009})} Given matrices $A,B \in \mathbb{R}^{n \times n }$:
\begin{equation*}
     \rho (A+B) \leq \sigma_{max}(A+B) \leq \sigma_{max}(A) + \sigma_{max}(B).   
\end{equation*}
    \label{Property_rho_sigma}
\end{property}
\begin{property} \textnormal{(Proposition 7.1.10., page 401 in \cite{Bernstein2009})}\\
    Let $A \in \mathbb{R}^{n \times n}$ and $B \in \mathbb{R}^{m \times m}$, then:
    \begin{equation*}
        \textnormal{mspec}(A \otimes B)= \{ \lambda \mu: \: \: \lambda \in \textnormal{mspec(A)}, \mu \in \textnormal{mspec}(B)\}_{ \textnormal{ms}}.
    \end{equation*}
    \label{Property_mspec_Kronecker}
\end{property}
The following theorem presents a sufficient condition allowing to design a pre-stabilizing gain ensuring $\rho(M(K)) < \alpha$, where $\alpha \in (0,1)$ denotes a contraction rate, implying therefore the convergence of the covariance dynamics, as mentioned in Corollary~\ref{Cor:Stability}.
\begin{theorem}\label{LMI_condition}
    Given $A_0 \in \mathbb{R}^{n \times n }$, $B\in \mathbb{R}^{n \times m}$ and $C_p \in \mathbb{R}^{n^2 \times n^2}$, if there exists $K \in \mathbb{R}^{m \times n}$ and $\alpha \in (0,1)$  such that the following holds: 
\begin{equation}
    \begin{bmatrix}
    (\alpha - \sigma_{max}(C_p)) I \ \  & (A_0+BK)^T\\
    (A_0+BK) & I 
    \end{bmatrix} \succ 0,
    \label{Eq:LMI_condition}
\end{equation}
then $K$ is such that $\rho(M(K) )<\alpha$.
\end{theorem}
\paragraph*{Proof}
Consider $A_K=A_0+BK$, using Property~\ref{Property_rho_sigma} on the matrix $A_K \otimes A_K + C_p $, we have:
\begin{equation}\label{Eq:conservative_bound1}
    \rho \left(A_K \otimes A_K + C_p  \right) \leq \sigma_{max} \left( A_K \otimes A_K\right) +\sigma_{max} \left(C_p\right).
\end{equation}
Therefore, $\sigma_{max} \left( A_K \otimes A_K\right) +\sigma_{max} \left(C_p\right) < \alpha$, with $\alpha \in (0,1)$, implies that $\rho \left(A_K \otimes A_K + C_p  \right) < \alpha$, then it is sufficient to impose that
\begin{equation}
    \sigma_{max} \left( A_K \otimes A_K\right) < \alpha - \sigma_{max} \left(C_p\right),
    \label{eq:conservative_bound}
\end{equation}
which,  since $\sigma_{max}(N)^2 = \lambda_{max}(N^TN)$, with $N$ a given real matrix, is also equivalent to:
\begin{equation}
    \lambda_{max} \left( \left( A_K^T \otimes A_K^T\right) \left( A_K \otimes A_K\right)\right)  < \left(\alpha - \sigma_{max} \left(C_p\right)\right)^2.
    \label{Eq:Lambda_max_Condition}
\end{equation}
By using Property~\ref{Property_Kronecker} and considering $\beta(\alpha ) = \alpha - \sigma_{max} \left(C_p\right)$, (\ref{Eq:Lambda_max_Condition}) is equivalent to: 
\begin{equation*}
        \lambda_{max} \left( \left( A_K^T A_K\right)  \otimes \left(A_K^T   A_K\right)\right)  < \beta(\alpha )^2,
\end{equation*}
which is equivalent to $\lambda_{max}^2 \left( A_K^T A_K\right)  < \beta(\alpha )^2$ (by using Property~\ref{Property_mspec_Kronecker}), and to $\lambda_{max} \left( A_K^T A_K\right)  < \beta(\alpha )$, that leads to $
A_K^T A_K \prec \beta(\alpha) I, 
$
which, by using the Schur complement, is equivalent to: 
\begin{equation*}
\begin{bmatrix} 
    \beta(\alpha ) I  & (A_0+BK)^T\\
(A_0+BK) & I 
\end{bmatrix} \succ 0.
\end{equation*}
\endproof
In practice, since $\beta(\alpha)$ is affine in $\alpha$,  the latter can be minimized  if the objective is not only to ensure stability but also to optimize the covariance convergence. Note that the condition provided by Theorem~\ref{LMI_condition} is conservative because of the successive bounds in (\ref{Eq:conservative_bound1}) and (\ref{eq:conservative_bound}).

As previously mentioned, the conditions provided in \cite{HOSOE2019} (Section~V) are necessary and sufficient for ensuring MSS of systems with multiplicative uncertainties, i.e. $x_{k+1} = A(p_k) x_k + B(p_k) u_k$, in closed-loop with a state feedback $u_k = K x_k$.
The dimension of the LMI conditions presented in \cite{HOSOE2019} is at most $(n^2(n+m)+n )\times (n^2(n+m)+n)$. This dimension can in fact be reduced; in particular, in the considered setting  where the input matrix is deterministic, it reduces to  $(n(n^2+2) )\times (n(n^2+2))$. Although the condition provided in Theorem~\ref{LMI_condition} is sufficient and more conservative, it offers the possibility of having a lower dimensional LMI condition ($2n \times 2n$) for a systematic design of $K$, which is more suitable for higher dimensional systems as will be illustrated subsequently. 

\section{Numerical example}
This  section presents numerical examples assessing the exact characterization of the error covariance dynamics and the numerical tractability of condition~(\ref{Eq:LMI_condition}).

Consider system~(\ref{Eq:sys_dyn}) with the following setting: 
\begin{equation}\label{Ex:sys1}
    A_0= 
        \begin{pmatrix}
0.4& 0.4 & 0.5 \\
0.1 & 0.1 & 0.2\\ 
0.2 & 0.1 & 0.5
\end{pmatrix},
\qquad
    \bar{A}(p_k)= 
        \begin{pmatrix}
p_{1k} & 0 & p_{1k} \\
0 & p_{2k}^2-0.15 & 0 \\
0 & 0 & p_{2k}
\end{pmatrix}.
\end{equation}
with $B = \left(1, 0, 0 \right)^T$,  $\mathbb{E}[w_kw_k^T]=I_{n}$ and $p_k = \left(p_{1k},p_{2k}\right)^T \: \mathtt{\sim} \: \mathcal{N}(0,\Sigma)$, where $\Sigma$ is defined as:
\begin{equation*}
\Sigma=
            \begin{pmatrix}
    0.21&    0.03\\
    0.03 & 0.15
\end{pmatrix}.
\end{equation*}
By solving the LMI condition~(\ref{Eq:LMI_condition}) and minimizing $\alpha$, we obtain $K=(-0.4\   -0.4\   -0.5)^T$ with $\alpha = 0.78$. In order to assess the conservatism of  condition~(\ref{Eq:LMI_condition}), we solve the LMI presented in Corollary~1 in \cite{HOSOE2019}, which gives $K=(  -0.76\   -0.59\   -1.39)^T$ with $\alpha = 0.34$, $\alpha$ here stands for $\lambda^2$ in \cite{HOSOE2019}, when no ambiguity may raise. This shows that the condition that we derive here is more conservative, we will assess later its practical usefulness from a computational point of view.

We compute the evolution of the vectorization of the error covariance using the difference equation in~(\ref{eq:err_cov_dynamics}), as well as the empirical covariance based on $N=5000$ trials. We denote by $\e_{ij}^{th}$ and $\e_{ij}^{em}$, respectively, the theoretical and the empirical  elements of the error covariance matrix, for $i,j\in \{1,2,3\}$, corresponding to states indices. Note that in this case we have $6$ non-redundant elements in the error covariance matrix. Fig.~\ref{fig:ex_sys1} shows that the empirical error covariance matches the theoretical one. This figure presents only $5$ trajectories, since in our example one of the entries is always null.  We can notice in Fig.~\ref{fig:ex_sys1} that both the empirical and theoretical covariance matrices converge to the following matrix (as stated in Corollary~\ref{Cor:Stability}):
\begin{equation*}
    \vec^{-1}\left(\left( I-M\right)^{-1} \vec \left(W\right)\right) =
    \begin{pmatrix}
    1.79     &    0  &  0.06\\
         0  &  0.17    &0.26\\
    0.06   & 0.26  &  1.87
\end{pmatrix}.
\label{Ex:cover_err}
\end{equation*}
Note that in this example, and for simulation purposes, the control $v_k$ is considered as a state feedback of the form $v_k = K z_k$, which is equivalent to considering $u_k = K x_k$. In an SMPC framework, $v_k$ is designed by a deterministic MPC strategy. 
\begin{figure}
    \centering 
\includegraphics[width=1\linewidth]{Cov-traj.tikz}
    \caption{Theoretical and empirical error covariance evolution related to system~(\ref{Eq:sys_dyn}) with setting~(\ref{Ex:sys1}).}
    \label{fig:ex_sys1}
\end{figure}

Finally, in order to assess the computational tractability of condition~(\ref{Eq:LMI_condition})  with $\alpha=1$ in comparison to condition~(59) in \cite{HOSOE2019}, we consider that only $A(p_k)$ is stochastic and the input matrix is deterministic. In this case, the size of the condition in \cite{HOSOE2019} is further reduced using the singular value decomposition as mentioned in \cite{Hosoe2019b}. Furthermore, we consider the following setting:
\begin{itemize}
    \item Random nominal matrices $A_0$ of size $n$ were generated such that both conditions were feasible and $B=I_n$. 
    \item The entries of $\bar{A}(p_k)$  are assumed to follow normal distributions $\mathcal{N}\left(0, 0.07\right)$.  
    \item The average computation time was evaluated using 100 tests, considering only the solver execution step, \textit{i.e.} without the decomposition required in \cite{HOSOE2019} and without decay rate minimization. 
\end{itemize} 
\begin{figure}
    \centering
\includegraphics[width=1\linewidth]{Fig_com_time.tikz}
    \caption{Average computation time over 100 tests}
    \label{fig:B_det}
\end{figure}
Fig~\ref{fig:B_det} shows the average computational time related to condition ~(\ref{Eq:LMI_condition}) is considerably lower than the one related to condition (59) in \cite{HOSOE2019}, which is expected from the dimension of both conditions. Therefore, although conservative, condition~(\ref{Eq:LMI_condition}) can be useful for higher dimensional systems since its dimension is linear in $n$. 

\section{Conclusion}
In this paper, we provide an SMPC-oriented characterization of the dynamics of the error covariance for discrete-time linear systems under potentially unbounded additive and parametric uncertainties and present an LMI-based condition for the stability of these dynamics. The characterization highlights the coupling between the nominal trajectory and the error covariance dynamics, which is absent in the standard additive-noise case and must be accounted for when designing reachable sets and constraint-tightening mechanisms in this context.  The proposed numerical example
shows that the theoretical and the empirical error covariance converge to the same matrix when the stability conditions are satisfied, and that the proposed control design condition is more tractable for higher dimensional systems. Future works would focus on using this characterization to design stochastic invariant sets and SMPC strategies.
\begin{ack}                               
This work was supported in part by the Clinical project, funded by the ANR under grant ANR-24-CE45-4255, in part by the FMJH Program Gaspard Monge for optimization and operations research and their interactions with data science.
\end{ack}

\bibliographystyle{plain}        
\bibliography{Paper/Biblio}

\begin{thebibliography}{10}

\bibitem{Arcari2023}
Elena Arcari, Andrea Iannelli, Andrea Carron, and Melanie~N. Zeilinger.
\newblock Stochastic {MPC} with robustness to bounded parameteric uncertainty.
\newblock {\em IEEE Transactions on Automatic Control}, 68(12):7601--7615,
  2023.

\bibitem{Bakolas2018}
Efstathios Bakolas.
\newblock Finite-horizon covariance control for discrete-time stochastic linear
  systems subject to input constraints.
\newblock {\em Automatica}, 91:61--68, 2018.

\bibitem{Bernstein2009}
Dennis~S. Bernstein.
\newblock {\em Matrix Mathematics, Theory, Facts, and Formulas (Second
  Edition)}.
\newblock Princeton University Press, Princeton, 2009.

\bibitem{Bertsekas2008}
Dimitri Bertsekas and John~N Tsitsiklis.
\newblock {\em Introduction to probability}, volume~1.
\newblock Athena Scientific, 2008.

\bibitem{Blackmore2010}
Lars Blackmore, Masahiro Ono, Askar Bektassov, and Brian~C. Williams.
\newblock A probabilistic particle-control approximation of chance-constrained
  stochastic predictive control.
\newblock {\em IEEE transactions on Robotics}, 26(3):502--517, 2010.

\bibitem{Boyd1994}
Stephen Boyd, Laurent El~Ghaoui, Eric Feron, and Venkataramanan Balakrishnan.
\newblock {\em Linear matrix inequalities in system and control theory}.
\newblock SIAM, 1994.

\bibitem{calafiore2012}
Giuseppe~C. Calafiore and Lorenzo Fagiano.
\newblock Robust model predictive control via scenario optimization.
\newblock {\em IEEE Transactions on Automatic Control}, 58(1):219--224, 2012.

\bibitem{Cannon2011}
Mark Cannon, Basil Kouvaritakis, Saša~V. Raković, and Qifeng Cheng.
\newblock Stochastic tubes in model predictive control with probabilistic
  constraints.
\newblock {\em IEEE Transactions on Automatic Control}, 56(1):194--200, 2011.

\bibitem{Collins1987}
Emman Collins and Robert~E. Skelton.
\newblock A theory of state covariance assignment for discrete systems.
\newblock {\em IEEE Transactions on Automatic Control}, 32(1):35--41, 1987.

\bibitem{FARINA2016}
Marcello Farina, Luca Giulioni, and Riccardo Scattolini.
\newblock Stochastic linear model predictive control with chance constraints
  – a review.
\newblock {\em Journal of Process Control}, 44:53--67, 2016.

\bibitem{Fiacchini2021}
Mirko Fiacchini and Teodoro Alamo.
\newblock Probabilistic reachable and invariant sets for linear systems with
  correlated disturbance.
\newblock {\em Automatica}, 132:109808, 2021.

\bibitem{Gravell2021}
Benjamin Gravell, Peyman~Mohajerin Esfahani, and Tyler Summers.
\newblock Learning optimal controllers for linear systems with multiplicative
  noise via policy gradient.
\newblock {\em IEEE Transactions on Automatic Control}, 66(11):5283--5298,
  2021.

\bibitem{Hewing2018}
Lukas Hewing and Melanie~N Zeilinger.
\newblock Stochastic model predictive control for linear systems using
  probabilistic reachable sets.
\newblock In {\em 2018 IEEE Conference on Decision and Control (CDC)}, pages
  5182--5188. IEEE, 2018.

\bibitem{HOSOE2019}
Yohei Hosoe and Tomomichi Hagiwara.
\newblock Equivalent stability notions, lyapunov inequality, and its
  application in discrete-time linear systems with stochastic dynamics
  determined by an i.i.d. process.
\newblock {\em IEEE Transactions on Automatic Control}, 64(11):4764--4771,
  2019.

\bibitem{Hosoe2019b}
Yohei Hosoe and Dimitri Peaucelle.
\newblock Static output feedback stabilization of discrete-time linear systems
  with stochastic dynamics determined by an independent identically distributed
  process.
\newblock {\em IEEE Control Systems Letters}, 3(3):673--678, 2019.

\bibitem{Hsieh1990}
Chen Hsieh and Robert~E Skelton.
\newblock All covariance controllers for linear discrete-time systems.
\newblock {\em IEEE Transactions on Automatic Control}, 35(8):908--915, 1990.

\bibitem{Knaup2023}
Jacob~W Knaup and Panagiotis Tsiotras.
\newblock Covariance steering for systems subject to unknown parameters.
\newblock In {\em 2023 62nd IEEE Conference on Decision and Control (CDC)},
  pages 1790--1795. IEEE, 2023.

\bibitem{KofmanAUT12}
Ernesto Kofman, José~A. {De Doná}, and Maria~M. Seron.
\newblock Probabilistic set invariance and ultimate boundedness.
\newblock {\em Automatica}, 48(10):2670--2676, 2012.

\bibitem{Langson}
Wilbur Langson, Ioannis Chryssochoos, Saša~V. Raković, and David~Q. Mayne.
\newblock Robust model predictive control using tubes.
\newblock {\em Automatica}, 40(1):125--133, 2004.

\bibitem{Li2005}
Weiwei Li, Emanuel Todorov, and Robert~E Skelton.
\newblock Estimation and control of systems with multiplicative noise via
  linear matrix inequalities.
\newblock In {\em Proceedings of the 2005, American Control Conference, 2005.},
  pages 1811--1816. IEEE, 2005.

\bibitem{Lorenzen2016}
Matthias Lorenzen, Fabrizio Dabbene, Roberto Tempo, and Frank Allg{\"o}wer.
\newblock Constraint-tightening and stability in stochastic model predictive
  control.
\newblock {\em IEEE Transactions on Automatic Control}, 62(7):3165--3177, 2016.

\bibitem{Okamoto2018}
Kazuhide Okamoto, Maxim Goldshtein, and Panagiotis Tsiotras.
\newblock Optimal covariance control for stochastic systems under chance
  constraints.
\newblock {\em IEEE Control Systems Letters}, 2(2):266--271, 2018.

\bibitem{Xing2020ACC}
Yu~Xing, Ben Gravell, Xingkang He, Karl~Henrik Johansson, and Tyler Summers.
\newblock Linear system identification under multiplicative noise from multiple
  trajectory data.
\newblock In {\em 2020 American Control Conference (ACC)}, pages 5157--5261.
  IEEE, 2020.

\bibitem{Xing2022AUT}
Yu~Xing, Benjamin Gravell, Xingkang He, Karl~Henrik Johansson, and Tyler~H
  Summers.
\newblock Identification of linear systems with multiplicative noise from
  multiple trajectory data.
\newblock {\em Automatica}, 144:110486, 2022.

\end{thebibliography}

\end{document}